\documentclass[final,psfig,epsfig,graphics]{aipproc}
\usepackage{graphicx}
\layoutstyle{6x9}
\def\be{\begin{equation}}
\def\ee{\end{equation}}
\def\bea{\begin{equnarry}}
\def\eea{\end{equnarry}}
\begin{document}
\title{Experimental Study of Dust Acoustic Waves in the Strongly Correlated Regime}
\author{\underline{P. Bandyopadhyay}}{
address={Institute for Plasma Research, Bhat, Gandhinagar-382428, India}}
\author{G. Prasad}{
address={Institute for Plasma Research, Bhat, Gandhinagar-382428, India}}
\author{A. Sen}{
address={Institute for Plasma Research, Bhat, Gandhinagar-382428, India}}
\author{P. K. Kaw}{
address={Institute for Plasma Research, Bhat, Gandhinagar-382428, India}}
\begin{abstract}
Low frequency dust acoustic waves (DAW) were excited in a laboratory argon dusty plasma by modulating the discharge voltage  with a low frequency AC signal. Metallic graphite particles were used as dust grains and a digital FFT technique was used to obtain dispersion characteristics. The experimental dispersion relation shows the reduction of phase velocity and a regime where $\partial \omega/\partial k < 0$. A comparison is made with existing theoretical  model.
\end{abstract}
\maketitle

\section{Introduction}
\indent  A dusty plasma with massive dust particles can experience significant  correlation effects when the Coulomb coupling  parameter $\Gamma>1$ (($\Gamma = \frac{Z_d^2e^2}{4\pi\epsilon_0dT_d}exp[-d/\lambda_p]$, where $Z_d$, $d$, $T_d$, $\lambda_p$ are number of charges on dust particles, interparticle distance, dust particle temperature and plasma debye length respectively). Past theoretical works \cite{carini,postogna,golden,berkov,kalman} suggest that correlation effects can significantly modify the collective modes of a dusty plasmas. Kaw and Sen \cite{sen} have studied the influence of correlation effect on DAW using a generalized hydrodynamic model and have obtained a new dispersive corrections, overall reduction in frequency and phase velocity, existence  of $\partial \omega /\partial k <0$ regime.\\
They have also pointed out that in most laboratory plasmas such correlation induced effects can get masked out by collisional  or ion streaming effects. To the best of our knowledge, such correlation induced effects have not been experimentally measured earlier. In the present work we report results on such a measurement and compare it to theoretical results.

\section{Experimental set-up}
\begin{figure}[h]
\centerline{\scalebox{0.3}{\includegraphics{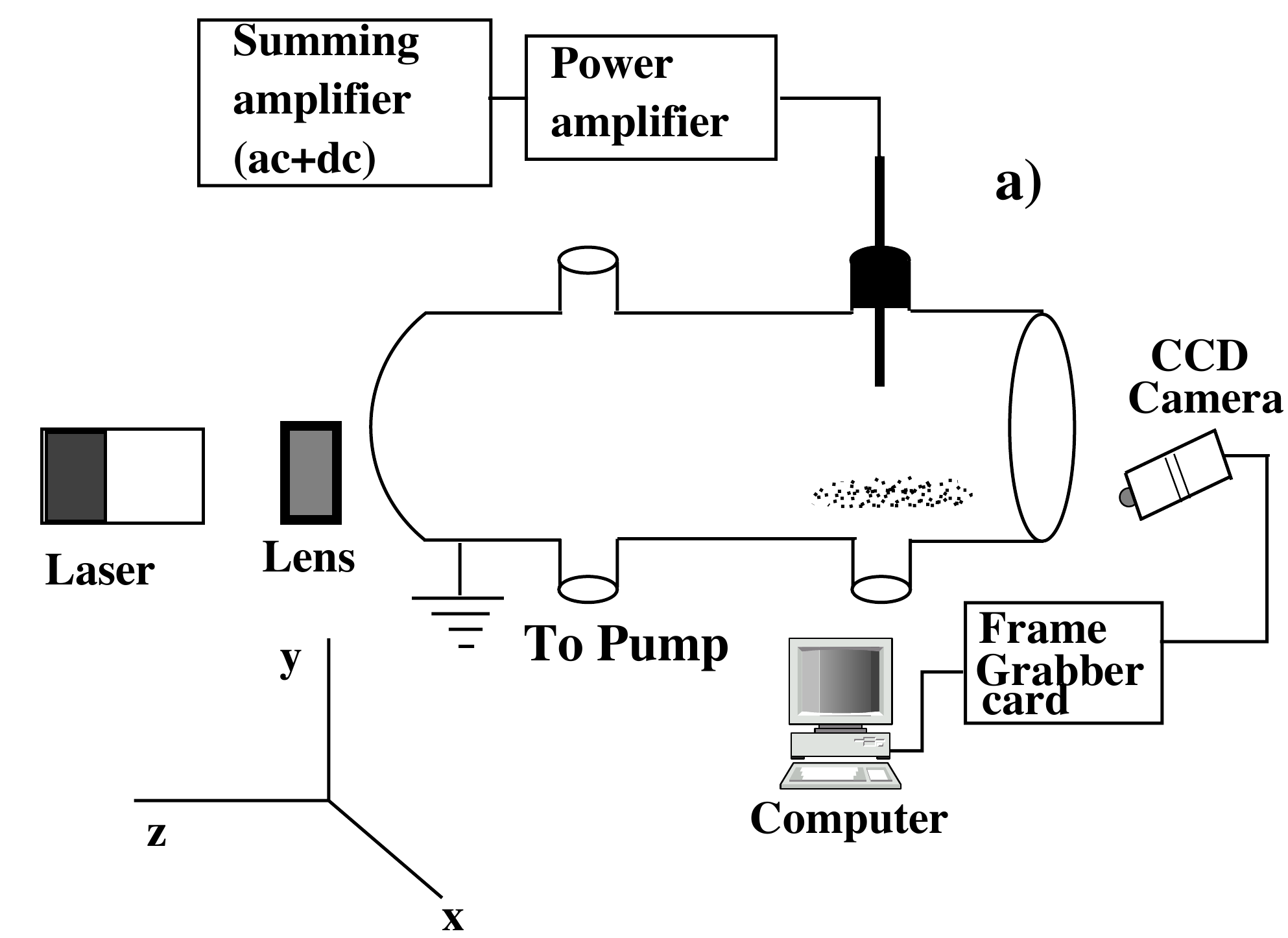}}
           \scalebox{0.35}{\includegraphics{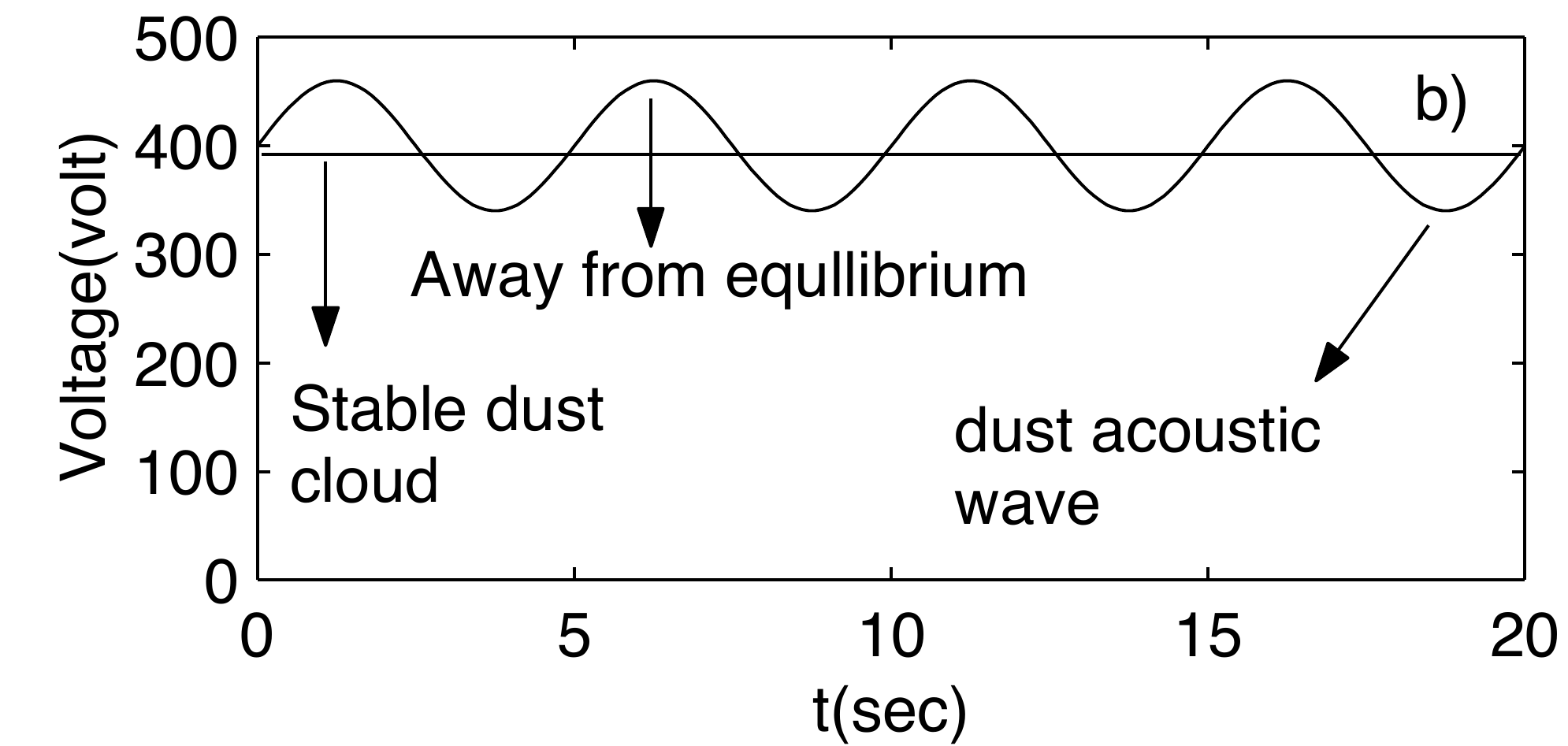}}}
\caption{a) Schematic of experimental setup and b) Typical discharge voltage waveform}
\label{fig:exp_setup}
\end{figure}

The schematic of  experimental setup is shown in Fig.\ref{fig:exp_setup}(a) and it is described in detail elsewere \cite{pramanik1}. The discharge was struck between the vacuum vessel (cathode) and rod shaped anode by apply a DC voltage ($V_{dc}$). The inner side of the cathode is covered with a thin stainless steel foil to avoid micro-arcs. Graphite flakes expanded in thermal plasma were used as dust in present experiment. These graphite  particles are of very low density ($\rho \sim 1.0 gm/cm^3$). The system was  pumped down to a base pressure ($P$) of $10^{-3}$mbar  using a rotary pump. It was filled with argon and plasma was formed by applying a discharge voltage ($V_{dc}$) $\sim500$ Volts to the anode at $P\sim1$ mbar. Then the pressure was reduced gradually to 0.086 mbar. It was observed that dust particles begin to accumulate near the anode region. The anode region provides suitable electric field against the gravitational field for levitation of dust grains. The radial electric field confines the dust particles.  An AC signal (using a signal generator and a power amplifier) was superimposed with the discharge voltage to excite the waves (See Fig.\ref{fig:exp_setup}(b)). The amplitude of the AC signal was increased with increase in pressure to excite the DAW. The frequency of applied AC signal was varied from 0 to 2 Hz very precisely. \\
The levitated dust particles were illuminated by green Nd-Yag diode laser light. The laser light was spread into a sheet by a cylindrical lens and forward scattered light from the dust cloud was used to visualize the dust particles. The scattered light from the dust particles was recorded using a CCD camera (25 fps) and later stored into a computer using a frame grabber card. The ion density ($n_i$), electron temperature ($T_e$) was measured using single a Langmuier probe. The dust temperature ($T_d$) was calculated from the velocity of the particles by tracing single particle in different frames. 
\begin{figure}[hbt]
\includegraphics[height=0.16\textheight]{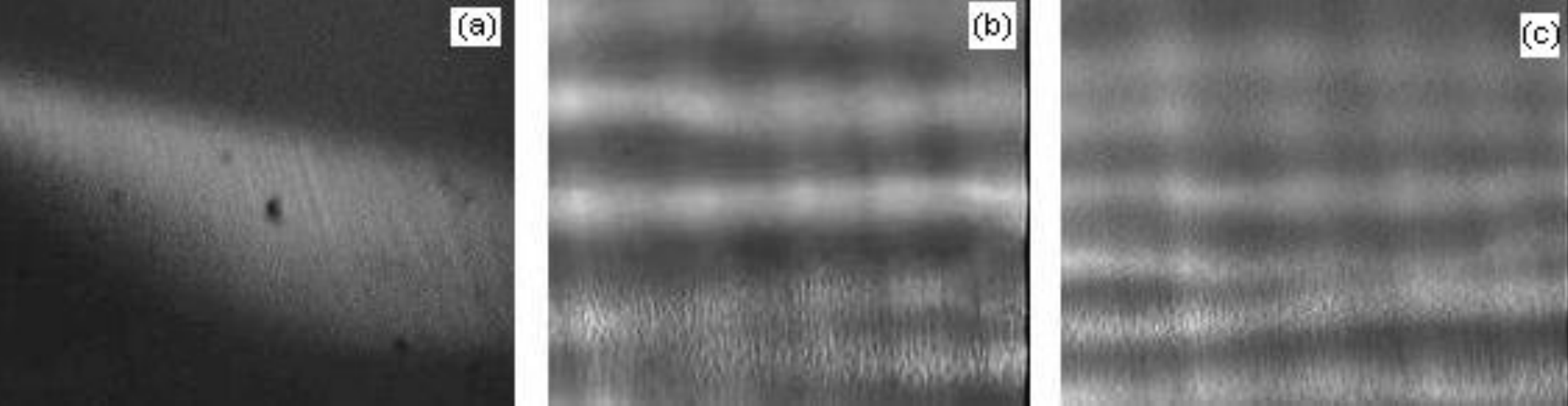}
\caption {a) Stable dust cloud b) $\&$ c) DAW of different wavelength}
\label{fig:wave}
\end{figure}
\section{Data Analysis}
Digitized data was arranged into several still images and stored as separate data records in the interval of 1/25 sec. Consecutive data records were then Fourier analyzed using the technique described by Smith \textit{et al} \cite{smith}, to obtain cross-power, phase and coherence ($\gamma$) spectra respectively. The frequency of the waves was computed using phase spectrum and time delay between consecutive frames.
\section{Result and Discussion}
The experiment was carried out to study the dispersive characteristics of dust acoustic wave in two different pressures regimes viz, low collisional regime (P=0.086 mbar) and high collisional regime (P=0.27 mbar) for a strongly coupled dusty plasma. In both regimes, initially a stable dust cloud was formed in the sheath region near the cathode (vessel wall). Typical stable dust cloud at above mentioned parameters is shown in the Fig. 2(a). It clearly shows that there is no background oscillations before applying AC signal. As the AC voltage was modulated with the DC voltage, the plasma glow as well as the discharge current exhibit oscillation of same frequency. The plasma boundary also varies with time, thus dust cloud as a whole also exhibits similar motion. In addition to this dust cloud breaks up into series of crests and troughs (see in Fig 2(b) and 2(c)) similar to pressure perturbation when the AC voltage  approaches minimum value. These perturbations have been referred in text as DAW oscillations and are similar to those reported earlier \cite{barkan,thompson,fortov,pramanik}. The dust cloud oscillates specially due to the application of oscillating voltage. A CCD camera was used to capture the images of dust cloud when it crossed the field of view. It was observed that DAW were excited just before the oscillating discharge voltage approaches minimum value. The frequency and wave length of the externally driven dust acoustic wave were measured from the FFT techniques of consecutive frames.
\begin{figure}[h]
\centerline{\scalebox{0.38}{\includegraphics{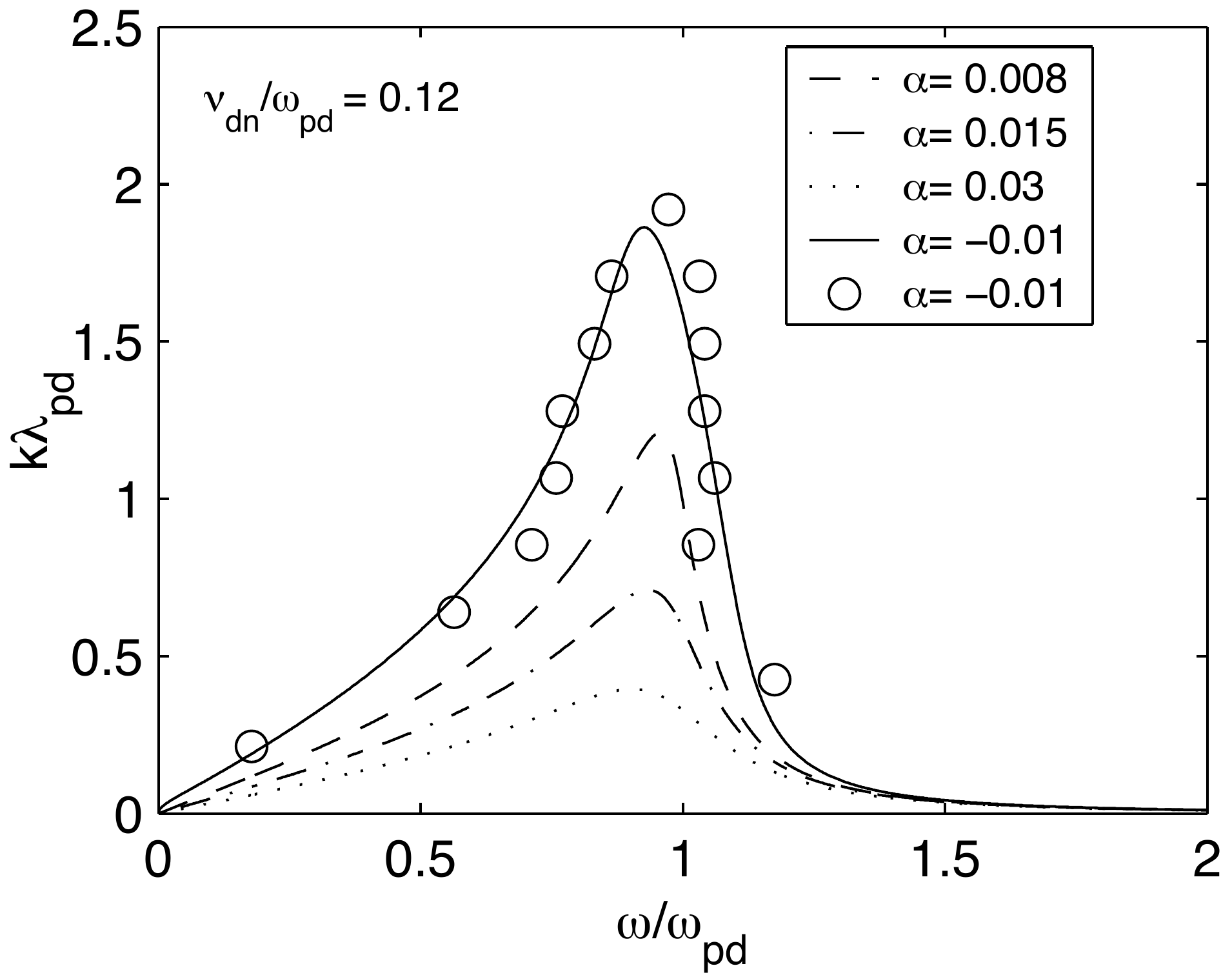}}
           \scalebox{0.38}{\includegraphics{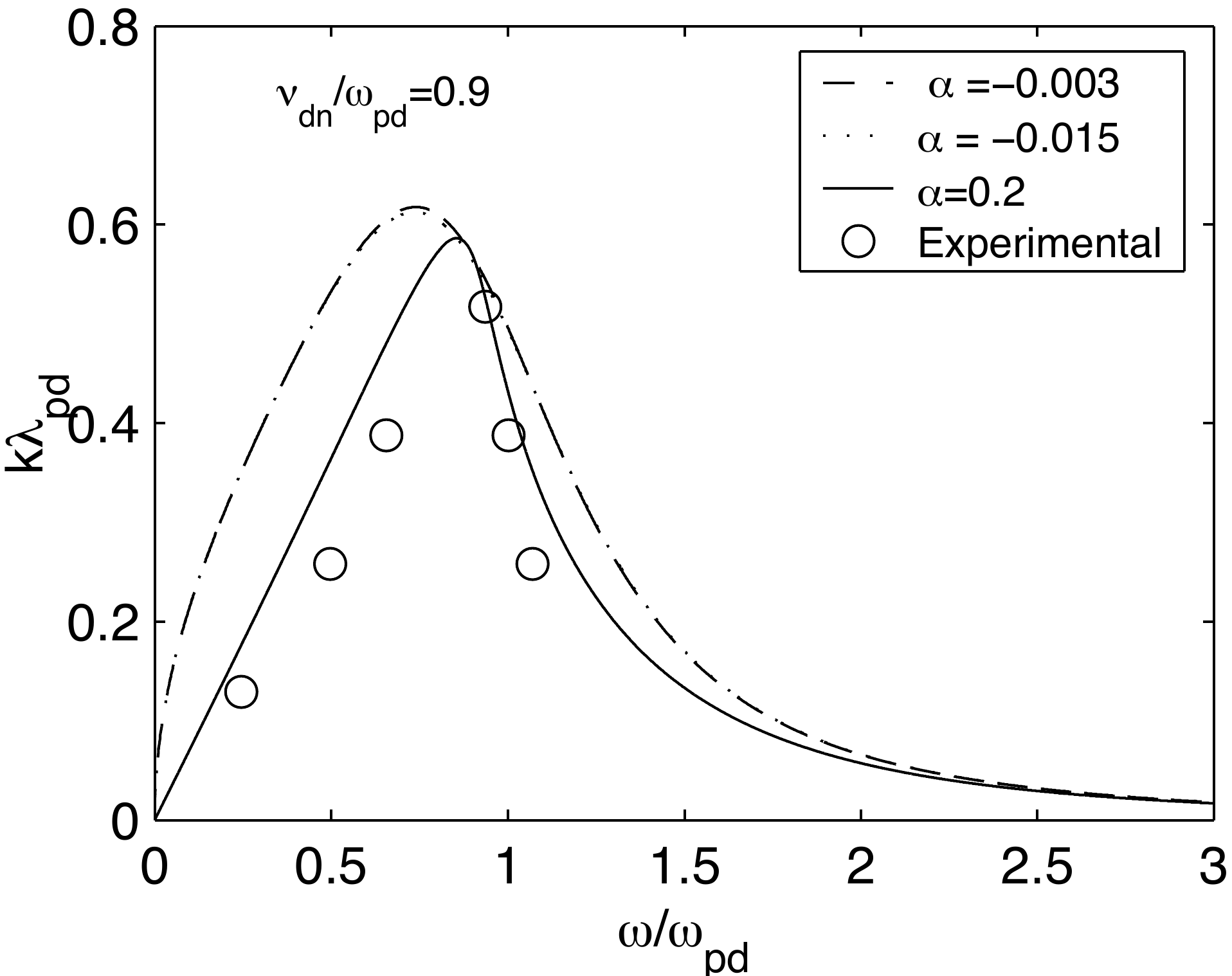}}}
\caption{Experimental (open circle) and theoretically calculated (solid and dotted line) dispersion curve for DAW in the a) low collisional regime and in the b) high collisional regime.}
\end{figure}
For our experimental parameter the phase velocity was quite low (few mm/sec) compare to other experimental groups \cite{thompson,pramanik,piper}. The dispersion relation of the externally driven DAW in both regime are shown in Fig 3(a) and 3(b). The open circles denote the experimental points. To compare our experimental results we choose a theory where strong correlation effects were taken into account along with the collisional effects \cite{sen}. We have plotted (with solid and dotted line) the dispersion relation according to the above theory for our experimental dusty plasma parameters. The DAW dispersion relation from the above theory is given by,
\begin{equation}
k^2\lambda_p^2  = \frac{\omega(\omega+i\nu_{dn}) - k^2 \beta}
{1-\omega(\omega+i\nu_{dn})+k^2\beta}
\end{equation}
Where $\beta = \gamma_d\mu_d \lambda_d^2$, $\gamma_d$ and $\mu_d$ are adiabatic index and coefficient of compressibility as mentioned in reference \cite{sen}. We have used experimentally measured $n_i$, $T_e$, $d$ and $T_d$ to calculate
$Z_d$, $\Gamma$, $\nu_{dn}$ and $\omega_{pd}=48$ was used as a free parameter to fit the theoretical dispersion curve \cite{piper}. The above dispersion relation is valid for hydrodynamic regime ($\omega_{pd} \tau_m \ll 1$, where $\tau_m$ is relaxation time).\\
As seen, there is a good agreement between the experimental and theoretical disperson curves for high and low collisional regime. In the strongly correlated regime $\alpha = \gamma_d \mu_d \lambda_d^2/\lambda_p^2 < 0$, since $\mu_d$ changes sign for $\Gamma > 3.1$ and $\alpha > 0$ in the weekly correlated regime  as discussed in \cite{sen}, the experimental dispersion relation shows negative slope ($\partial\omega/\partial k <0$) which is consistent with the theoretical model results predicted by Kaw et al. It is clear from the figures, at low collision frequencies the dispersion curves for strongly and weakly correlated regime are very different where as at higher collisional frequencies the differences are not significant. This is because the modifications of the dispersion relation due to dust neutral collisions are much stronger than correlation effects in higher pressure. Close match of theoretical prediction with experimental observation clearly suggest that correlation effect is significant in our case. Substantial reduction of peak value of dispersive curve in figure 3(b) shows the effect of strong collision.\\
\section{conclusion}
We have carried out an experiment on externally excited dust acoustic wave in a strongly coupled regime. The measured dispersive characteristic shows substantial reduction in phase velocity and existence of $\partial \omega/\partial k < 0$ regime. The dispersion relation at low pressure shows the effect of correlation. The correlation effects gets masked by collisional effect at high neutral pressure. These observations appear with past theoretical predictions.


\begin{thebibliography}{1}
\bibitem{carini}
P. Carini, G. Kalman and K. I. Golden, Phys. Rev. A. {\bf 26}, 1686, (1982).
P. Kairani and G. Kalman, Phys. Lett. A {\bf 105}, 232 (1984).
\bibitem{postogna}
F. Postogna and M. P. Tosi, Nuovo Cimento B{\bf 55}, 399 (1980).
\bibitem{golden}
K. I. Golden, Phys. Rev. A {\bf 35}, 5278 (1987).
\bibitem{berkov}
M. A.  Bervosky, Phys. Lett. A {\bf 166}, 365 (1992).
\bibitem{kalman}
G. Kalman, M. Rosenbarg and H. E. DeWitt, Phys. Rev. Lett., {\bf 84}, 6030, 2000.
\bibitem{sen}
P. K. Kaw, A. Sen, Phys. Plasmas, {\bf 5}, 3552 (1998).
\bibitem{pramanik1}
J. Pramanik, G. Prasad, A. Sen, P. K. Kaw, Phys. Rev. Lett. {\bf 88}, 175001, (2002).
\bibitem{smith}
D.E Smith, E.J. Powers and G.S.Caldwell, IEEE Transaction on Plasma Science,
{\bf PS-2}, 261 (1974).
\bibitem{barkan}
A. Barkan, R.L. Marlino, N. D'Angelo, Phys. Plasmas, {\bf 2}, 3563 (1995).
\bibitem{thompson}
C. Thompson, A. Barkan, N. D'Angelo, R.L. Marlino, Phys. Plasmas {\bf 7}, 1374 (2000). 
\bibitem{fortov}
V.E. Fortov et al., Phys. Plasmas, {\bf 7}, 1374, 2000.
\bibitem{pramanik}
J. Pramanik, B. M Veeresha, G. Prasad, A.Sen, P.K. Kaw, Phys Lett A, {\bf 312}, 84, (2003).
\bibitem{piper}
J.B Piper, J. Goree, Phys Rev. Lett., {\bf 77},3137 (1996).
\end{thebibliography}
\end{document}